\newcommand{\CLASSINPUTtoptextmargin}{0.77in}
\newcommand{\CLASSINPUTbottomtextmargin}{1.00in}
\documentclass[conference]{IEEEtran}
\IEEEoverridecommandlockouts
\usepackage{cite}
\usepackage{amsmath,amssymb,amsfonts}
\usepackage{algorithmic}
\usepackage{graphicx}
\usepackage{float}
\usepackage{textcomp}
\usepackage{glossaries}
\usepackage{xcolor}
\usepackage{array}
\usepackage{colortbl}
\usepackage{caption}
\usepackage{url}
\usepackage{hyperref}

\def\BibTeX{{\rm B\kern-.05em{\sc i\kern-.025em b}\kern-.08em
    T\kern-.1667em\lower.7ex\hbox{E}\kern-.125emX}}

\definecolor{lightgray}{gray}{0.9}
\definecolor{lightblue}{rgb}{0.678, 0.847, 0.902}

\newacronym{can}{CAN}{Controller Area Network}
\newacronym{bev}{BEV}{Battery Electric Vehicle} 
\newacronym{ecu}{ECU}{Electric Control Unit}
\newacronym{ids}{IDS}{Intrusion Detection System}
\newacronym{hcrl}{HCRL}{Hacking and Countermeasure Research Lab}
\newacronym{crysis}{CrySyS Lab}{Laboratory of Cryptography and System Security}
\newacronym{v2v}{V2V}{Vehicle-to-Vehicle}
\newacronym{lstm}{LSTM}{Long Short-Term Memory}
\newacronym{acc}{ACC}{Adaptive Cruise Control}
\newacronym{cacc}{CACC}{Cooperative ACC}
\newacronym{dbc}{DBC}{CAN DataBase}
\newacronym{suv}{SUV}{Sport Utility Vehicle}
\begin{document}

\title{CARACAS: vehiCular ArchitectuRe for detAiled Can Attacks Simulation
\thanks{
    This study was carried out within the SERICS - Security and Rights in the CyberSpace and received funding from the European Union Next-GenerationEU (PIANO NAZIONALE DI RIPRESA E RESILIENZA (PNRR) – MISSIONE 4 COMPONENTE 2, INVESTIMENTO 1.3 – D.D. 1556 11/10/2022, PE00000014), and within the COLTRANE-V project – funded by the Ministero dell'Università e della Ricerca – within the PRIN 2022 program (D.D.104 - 02/02/2022). This manuscript reflects only the authors' views and opinions. Neither the European Union, nor the European Commission, nor the Ministry can be considered responsible for them.\\
    (*) Correspondence: alessandro.savino@polito.it
  }
}

\author{\IEEEauthorblockN{Sadek Misto Kirdi$^{1}$, Nicola Scarano$^1$, Franco Oberti$^2$, Luca Mannella$^1$, Stefano Di Carlo$^1$, Alessandro Savino$^1*$}
\IEEEauthorblockA{
$^1$\textit{Politecnico di Torino, Department of Control and Computer Engineering, Turin, Italy}\\
$^2$\textit{Dumarey Softronix S.r.l., Torino}\\
\textit{e-mail: $^1$\{name.surname\}@polito.it, sadek.mistokirdi@studenti.polito.it,$^2$franco.oberti@dumarey.com}
}
}

\maketitle

\begin{abstract}
Modern vehicles are increasingly vulnerable to attacks that exploit network infrastructures, particularly the \gls{can} networks. To effectively counter such threats using contemporary tools like \glspl{ids} based on data analysis and classification, large datasets of \gls{can} messages become imperative.

This paper delves into the feasibility of generating synthetic datasets by harnessing the modeling capabilities of simulation frameworks such as Simulink coupled with a robust representation of attack models to present CARACAS, a vehicular model, including component control via \gls{can} messages and attack injection capabilities. CARACAS showcases the efficacy of this methodology, including a \gls{bev} model, and focuses on attacks targeting torque control in two distinct scenarios.
\end{abstract}


\glsresetall

\section{Introduction}



The automotive industry is undergoing a substantial shift in its primary drivers of innovation, moving from combustion engines to \glspl{bev}. This transition into a relatively unexplored domain of vehicular systems introduces unprecedented risks that car manufacturers have not previously encountered~\cite{Kim:2020aa}. Among them, cyberattacks are becoming more frequent, posing significant threats to both the systems' security and the drivers' safety~\cite{Upstream:aa}.

The \gls{can}~\cite{Szydlowski:1992aa}, which serves as the communication backbone among various components within traditional combustion engine vehicles, continues to be the primary communication technology in future cars. It is essential for enabling communication among multiple \glspl{ecu}, sensors, and intelligent actuators. This facilitates more straightforward vehicle wiring and sensor integration. However, the \gls{can} network's external accessibility provides attackers with multiple exploitation opportunities~\cite{Avatefipour:2017aa,Oberti:2021aa,Oberti:2024aa}. 
Furthermore, emerging functions such as \gls{acc} and \gls{cacc} systems, which are enabled by \gls{v2v} communication, have significantly enhanced vehicle efficiency and safety. Conversely, these systems expand the vehicle's attack surface and introduce new security vulnerabilities. Once an adversary gains access to a vehicle's \gls{can}, they can manipulate the system by injecting malicious packets. For instance, they could shut off the engine~\cite{Bi:2023aa} or turn off the immobilizer through a replay attack—a method currently popular for stealing luxury \glspl{suv}~\cite{stelvio}. These vulnerabilities require urgent attention and mitigation~\cite{Jo:2022aa}.

\glspl{ids} are countermeasures of great interest due to their simplicity and efficiency in detecting attacks, even on resource-constrained environments~\cite{Han:2018aa, IoTProxy2024, Chenet:2024aa}. They monitor network or host activities, raise alarms when unexpected events occur, and help prevent unauthorized access to the system. These techniques are being used extensively in Automotive, and multiple studies on the application of \gls{ids} on \gls{can} attacks have emerged, showing interesting results~\cite{Song:2016aa, Lokman:2019aa, Seo:2019aa}. 
Research in \gls{can} \gls{ids} has expanded quickly but needs help with reproducing, replicating, and comparing methodologies due to the need for costly infrastructure and extensive expertise~\cite{Lee:2017aa}. Consequently, due to their unavailability, many proposed detection techniques still need to be tested on suitable data~\cite{Verma:2020aa}.


The paper presents two main contributions. Firstly, it introduces a novel modular approach integrating a vehicular dynamics simulation model in a Simulink environment alongside a \gls{can} bus model and a \gls{can} injection model. It leverages the Simscape~\cite{MATLAB:2024aa} framework, a platform for building and simulating physical systems within Simulink~\cite{Documentation:2020aa} to simulate \gls{can} bus attacks.
This integration streamlines the analysis and validation of simulated \gls{can} attacks on the modeled vehicle, facilitating the generation of synthetic \gls{can} traffic and simulated attack data, allowing researchers to directly observe the effects of cyber threats on the vehicle's operational integrity and safety. 
Secondly, the paper demonstrates the capability of generating regular \gls{can} messages alongside malicious \gls{can} ones in a Simulink model for a \glspl{bev} vehicle based on the Tesla Model 3.
These models undergo testing across two driving scenarios to assess their performance when exposed to \gls{can} bus attacks, i.e., Extra Urban Driving Cycle and Cruise Mode.
The final generated dataset contains regular \gls{can} traffic and malicious torque \gls{can} messages.

The remaining paper includes Section~\ref{sec:related_works}, which analyzes the state of the art in modeling and dataset generation of \gls{can} attacks. In contrast, Section~\ref{sec:proposed_apporach} describes the modeling approach. Eventually, Section~\ref{sec:results} presents the simulated attacks results, and Section~\ref{sec:conclusion} sets the conclusion and future work perspectives.

\begin{figure*}[tb]
    \centering
    \includegraphics[width=0.90\textwidth]{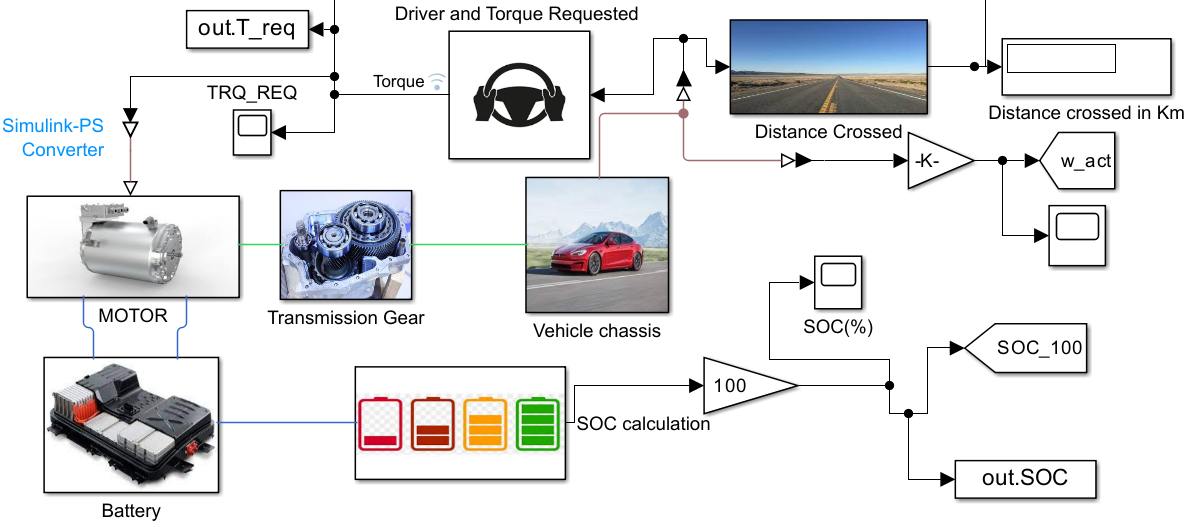}
    \caption{Full \gls{bev} Simulink model.}
    \label{fig:bev}
\end{figure*}


\section{Related Works}
\label{sec:related_works}

The rapid expansion of research in \gls{can} \glspl{ids} often encounters impediments due to challenges in replicating and comparing methodologies, primarily caused by difficulties obtaining valuable data~\cite{Hoppe:2011aa}. As a result, many detection techniques proposed still need to be validated due to the constrained availability of appropriate datasets~\cite{Verma:2020aa}. Over time, numerous datasets have emerged to alleviate data scarcity concerns. These datasets encompass real-world scenarios involving vehicles subjected to network attacks. For instance, \gls{hcrl} has released three datasets, each varying in data types, vehicles, and attack scenarios~\cite{Lee:2017aa, Han:2018aa, Seo:2019aa}. Additionally, the \gls{crysis} team at Budapest University of Technology and Economics has published the CrySyS dataset comprising \gls{can} traffic logs~\cite{Gazdag:2023aa}, while Eindhoven University of Technology has contributed datasets that combine real \gls{can} data with simulated attack injections~\cite{Dupont:2019aa}.

However, these real-world datasets show significant challenges during the data collection procedure, which is time-consuming and expensive, thus limiting the number of vehicles and the range of attacks used. A potential remedy for these limitations is the creation of simulated environments to generate artificial \gls{can} signals and simulate attacks within. For instance, Hanselmann et al. from Bosch GmbH, in~\cite{Hanselmann:2020aa}, developed a synthetic \gls{can} dataset to train and test their \gls{lstm}-based anomaly detector, ``CANet''. Despite its utility, the main drawback with this and other synthetic datasets (like the synthetic data from TU Eindhoven~\cite{Dupont:2019aa}) is the lack of verification of their effects on actual vehicles~\cite{Verma:2020aa}, posing doubts about the consequences of these attacks on vehicle systems.

The framework presented in this paper is designed to generate \gls{can} vehicle traffic under normal operating conditions by emulating a closed-loop model that adjusts based on critical system parameters. This framework facilitates the easy integration of various car models, enhancing its applicability across different automotive technologies.
Furthermore, our \gls{can} Injector module is designed to emulate a broad spectrum of external hacking tools used by attackers \cite{hacktool}. This enhancement significantly increases the adaptability and robustness of our testing process, thereby making it more effective in identifying vulnerabilities.

Ultimately, the capabilities of the injector are crucial, particularly in determining the severity of attacks and their correlation with the system's anticipated behavior on various road conditions.


\section{Proposed Approach}
\label{sec:proposed_apporach}

The proposed system is composed of three interconnected modules designed using Simulink~\cite{Documentation:2020aa}: (i) a \gls{bev} Dynamic model, (ii) a \gls{can} bus model, and (iii) a \gls{can} Attack Injection simulator.

\begin{figure*}[tb]
    \centering
    \includegraphics[width=0.99\textwidth]{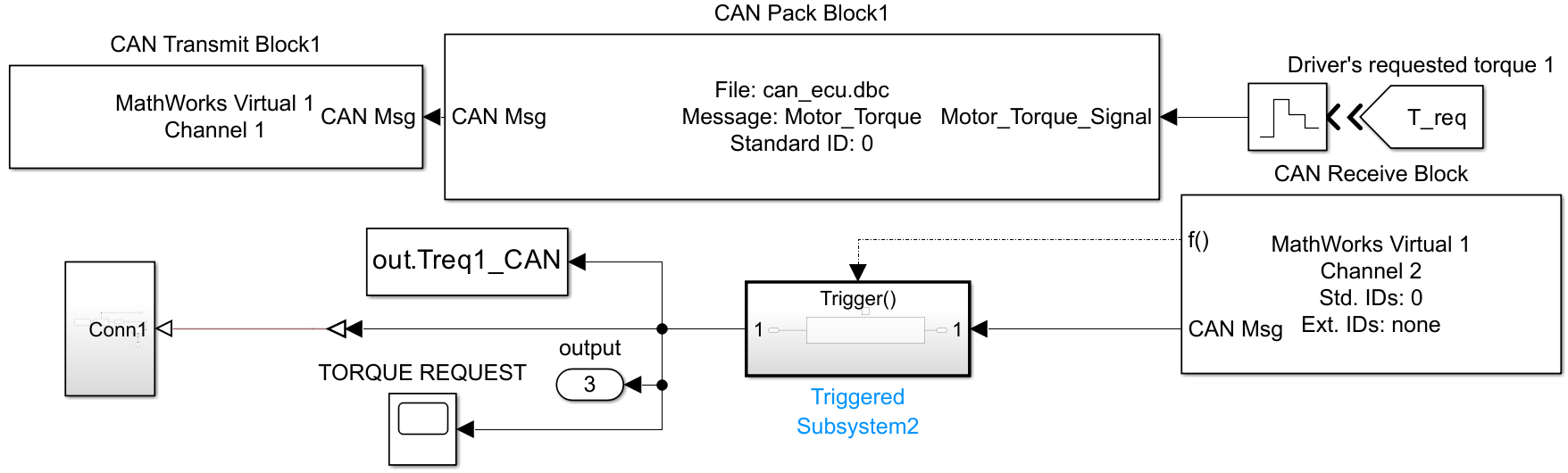}
    \caption{Torque request schema through \gls{can}.}
    \label{fig:can_config}
\end{figure*}

\subsection{BEV Dynamic model}

The \glspl{bev} behaviors on driving scenarios are supported by a Simscape~\cite{MATLAB:2024aa} model built within the Simulink environment. The \gls{bev} model, shown in Figure~\ref{fig:bev}, integrates the essential components of a vehicle system, such as the motor, chassis, battery, and control algorithms. These interconnected components were designed to emulate not only the core dynamics of an electric vehicle but also \gls{can} message transmissions to allow the assessment of these \gls{can} attacks' impacts.

The chassis serves as the vehicle's structural backbone, incorporating vital features like physical attributes and weight distribution, which are crucial for accurately simulating vehicle dynamics. A simplified representation of the battery was employed to power the motor and showcase the dynamics of the battery charge, sidestepping the complexities typical in real-world battery management systems. The motor was modeled with a focus on its performance characteristics and how it interacts with the rest of the vehicle's powertrain. A vital parameter of the motor (used as the target of the simulated attacks in this work) is torque, which represents the amount of rotational force the motor develops. It is responsible for the acceleration and breaking of a vehicle.

This combination of components within the simulation framework supports validating vehicle behaviors under various operational conditions. It enhances the ability to adjust and scale the model for future, more complex scenarios. 

\subsection{CAN bus model integration}

The CANdb++ Editor is a software tool from Vector~\cite{Vector-Informatik-GmbH:2024aa} essential in automotive and industrial applications for designing, editing, and managing \gls{can} communication protocol databases. Its capacity to precisely define message frames and signal layouts makes it invaluable for creating controlled synthetic attack scenarios on the \gls{can} network. Using the CANdb++ Editor, users can rigorously set up and modify a database, which is necessary for simulating specific network behaviors, including malicious attacks. This tool enables crafting unique messages and signal configurations that mimic potential security threats, thus providing a realistic environment for testing the robustness of \gls{bev} against \gls{can} injections and other related cyber-attacks.

By configuring the specific attributes of the \gls{can} messages and defining how these messages interact with various \glspl{ecu} in the network, engineers can simulate and monitor how these synthetic attacks would affect the vehicle's operation. This includes observing changes in vehicle parameters such as torque, speed, and state of charge under attack scenarios.

We set the CANdb++ Editor with the parameter needed for our simulation and generated the \gls{dbc} file.
Several Simulink Blocks contribute to integrating the \gls{can} into the model. 
%
The designed system and the interactions among the blocks are graphically represented in Figure~\ref{fig:can_config}. The \gls{can} Configuration block allows selecting the device, channel, and bus speed. Once the signals are configured, a Simulink Zero-Order-Hold block converts continuous signals to discrete ones suitable for \gls{can} communication. Following this, the \gls{can} Pack block encapsulates these signals into \gls{can} messages, which are subsequently transmitted over the network using the \gls{can} Transmit block. The \gls{can} Receive block captures these messages and delivers them to the Simulink vehicle model through the Trigger Block. In the latter, a \gls{can} Unpack block converts the \gls{can} data into signals for the Simulink model.

\subsection{CAN Attack Injection simulator}

\begin{figure*}[tb]
    \centering
    \includegraphics[width=0.99\textwidth]{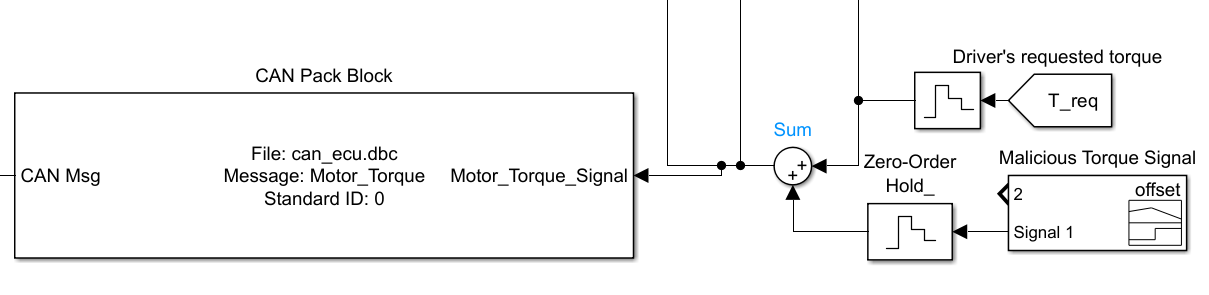}
    \caption{Torque attack injector integration into the \gls{can} model.}
    \label{fig:injector}
\end{figure*}


To conduct the simulation of a \gls{can} injection attack, we utilized the Simulink signal builder block (Figure~\ref{fig:injector}). The signal builder enabled us to construct various attack signals to replicate malicious scenarios and observe their impact on the vehicle's functionalities. Eventually, the signal builder provides a fast way to craft the related \gls{can} messages that perform the attack.

Figure~\ref{fig:injector} shows how the attack, thanks to a summing block, can be easily programmed separately, leaving the original model unaltered. Such modularity also allows for the setting up of different attack models within the same experimental setup. Moreover, the trigger-based activation of the attack simplifies the data collection of benign and malicious messages in the same data generation campaign. As the manipulation is at the signal level, it is easy for the user to manipulate the frequency and amplitude of the signals to vary the attack's modality instead of crafting \gls{can} messages directly.

The \textit{attack scenarios} simulated in this work involve a torque attack on the \gls{can} bus during the Extra Urban Driving Cycle and while in Cruise Mode. This attack injects malicious torque signals into the network, altering the average motor torque and consequently affecting the vehicle's speed.

\section{Results}
\label{sec:results}

\begin{figure}[h]
    \centering
    \includegraphics[width=0.49\textwidth]{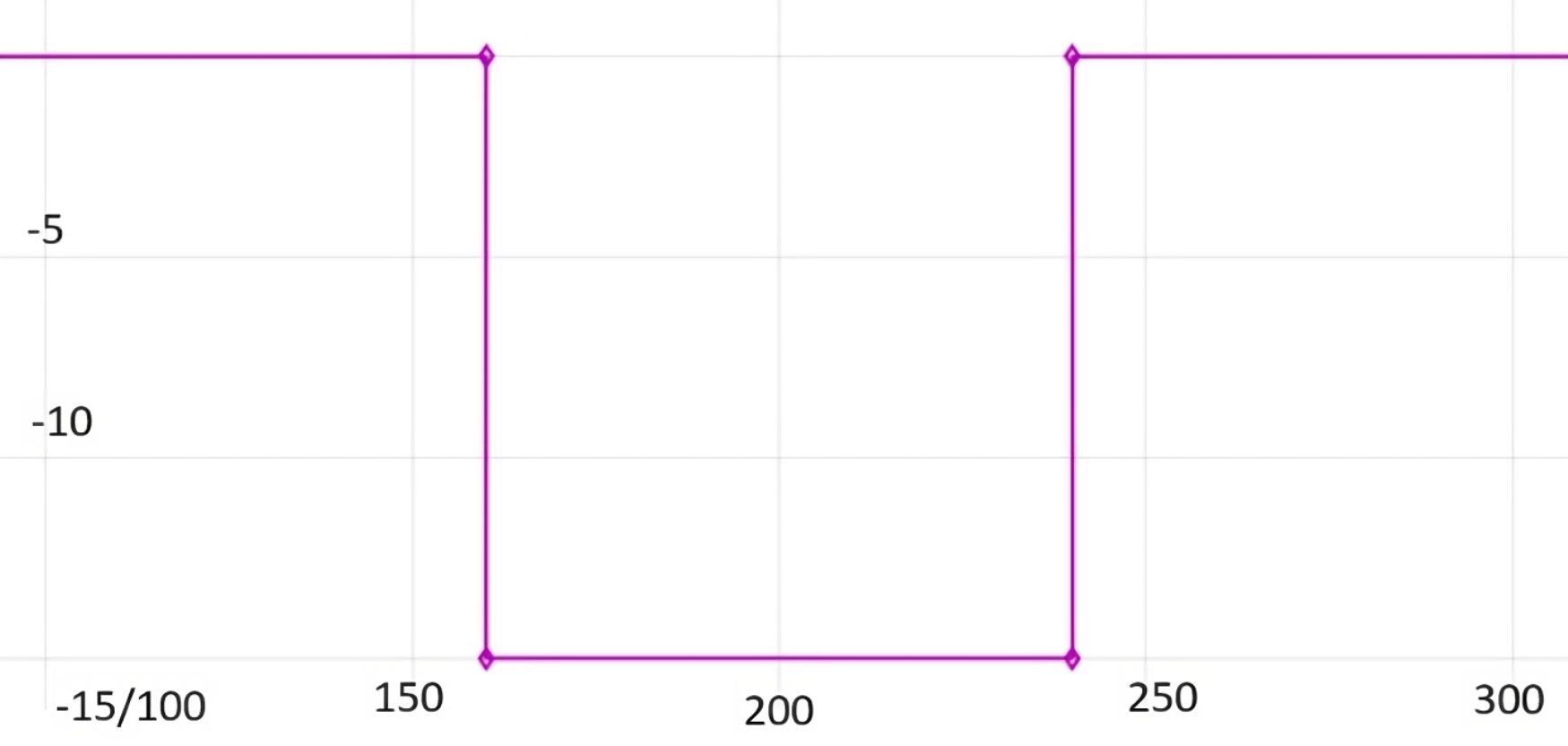}
    \caption{Step signal injected to simulate a braking torque. Simulation time in seconds [s] on the x-axis and torque in Newton meters [Nm] on the y-axis.}
    \label{fig:step_zoomed}
\end{figure}

To assess the potentiality of the modeling approach, the evaluation involved simulating a \gls{can} attack on a Tesla Model 3 during an Extra Urban Driving Cycle (with varying velocity and torque values) and Cruise Mode at constant velocity. In both scenarios, the injection of a braking torque signal of $-15nm$ is strategically timed to occur between $t = 160s$ and $t = 240s$ (Figure ~\ref{fig:step_zoomed}). 

The aim is to simulate a physical attack, then check if the model adheres to the realistic consequences, demonstrating the correct modeling, and if the \gls{can} messages follow, proving that a valid dataset can be generated.

\subsection{Extra Urban Driving Cycle} \label{subsec:EUDC}

\begin{figure*}[tb]
    \centering
    \includegraphics[width=0.70\textwidth]{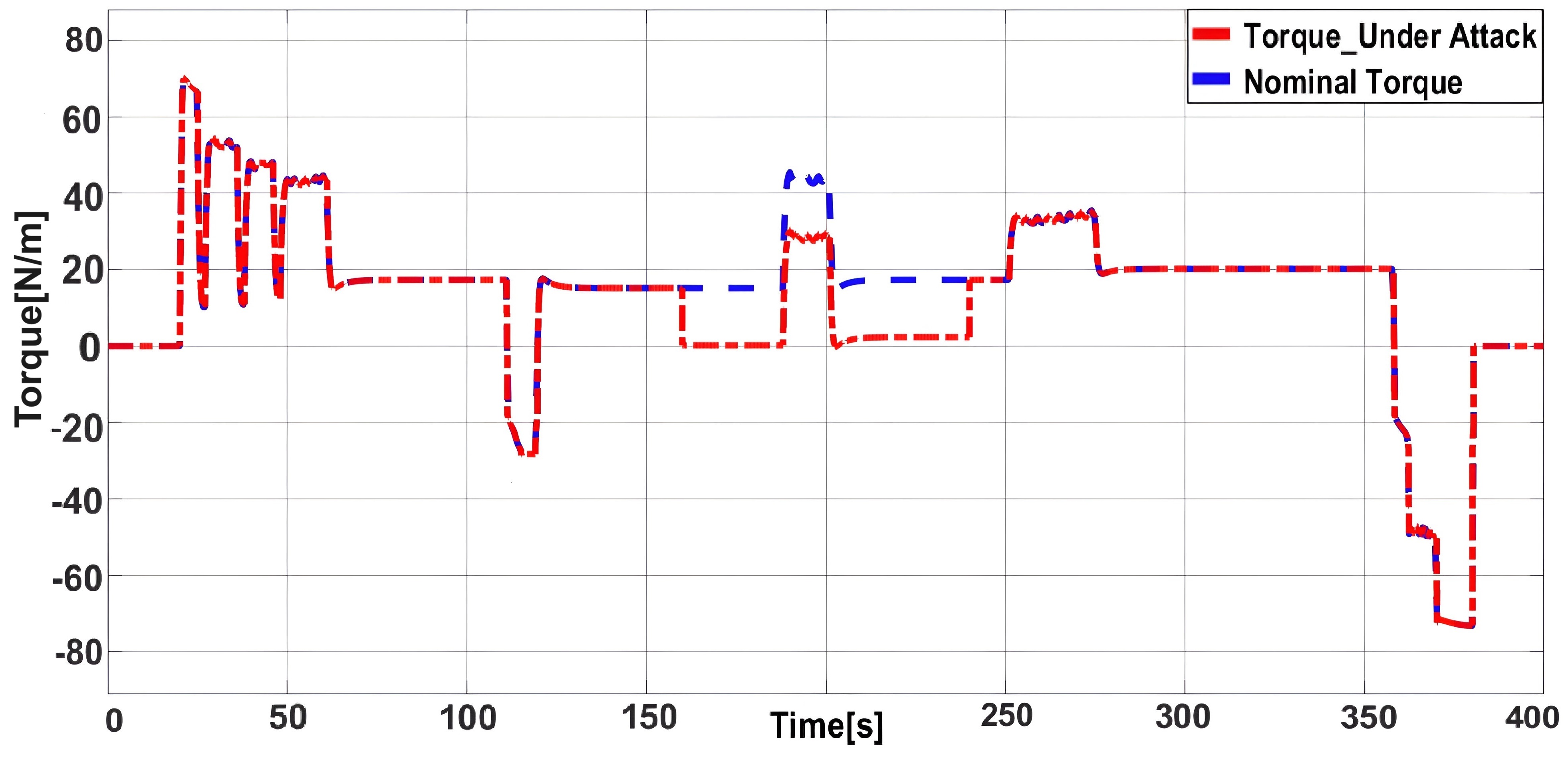}
    \caption{Torque without any attack vs Torque with the attack while the vehicle is in Extra Urban Driving Cycle. 
    }
    \label{fig:torque}
\end{figure*}

Figure~\ref{fig:torque} highlights the comparative effects of the torque with and without the attack. Before $t = 160s$, both the not attacked (shown in blue) and attacked (shown in red) torque signals displayed identical values, as no attack was implemented during this initial phase.
However, at $t=160 s$, a discernible divergence occurs between the two signals. The regular torque signal, i.e., the one which is not attacked, remains stable at approximately $15 Nm$. In contrast, when the attack started, the torque signal visibly dropped by an equivalent magnitude of $15 Nm$ below the baseline, effectively mirroring the parameters of the engineered attack signal. This shift clearly illustrates the direct impact of the \gls{can} intrusion, manifesting as an altered torque output that deviates from the expected performance.

\begin{figure*}[tb]
    \centering
    \includegraphics[width=0.70\textwidth]{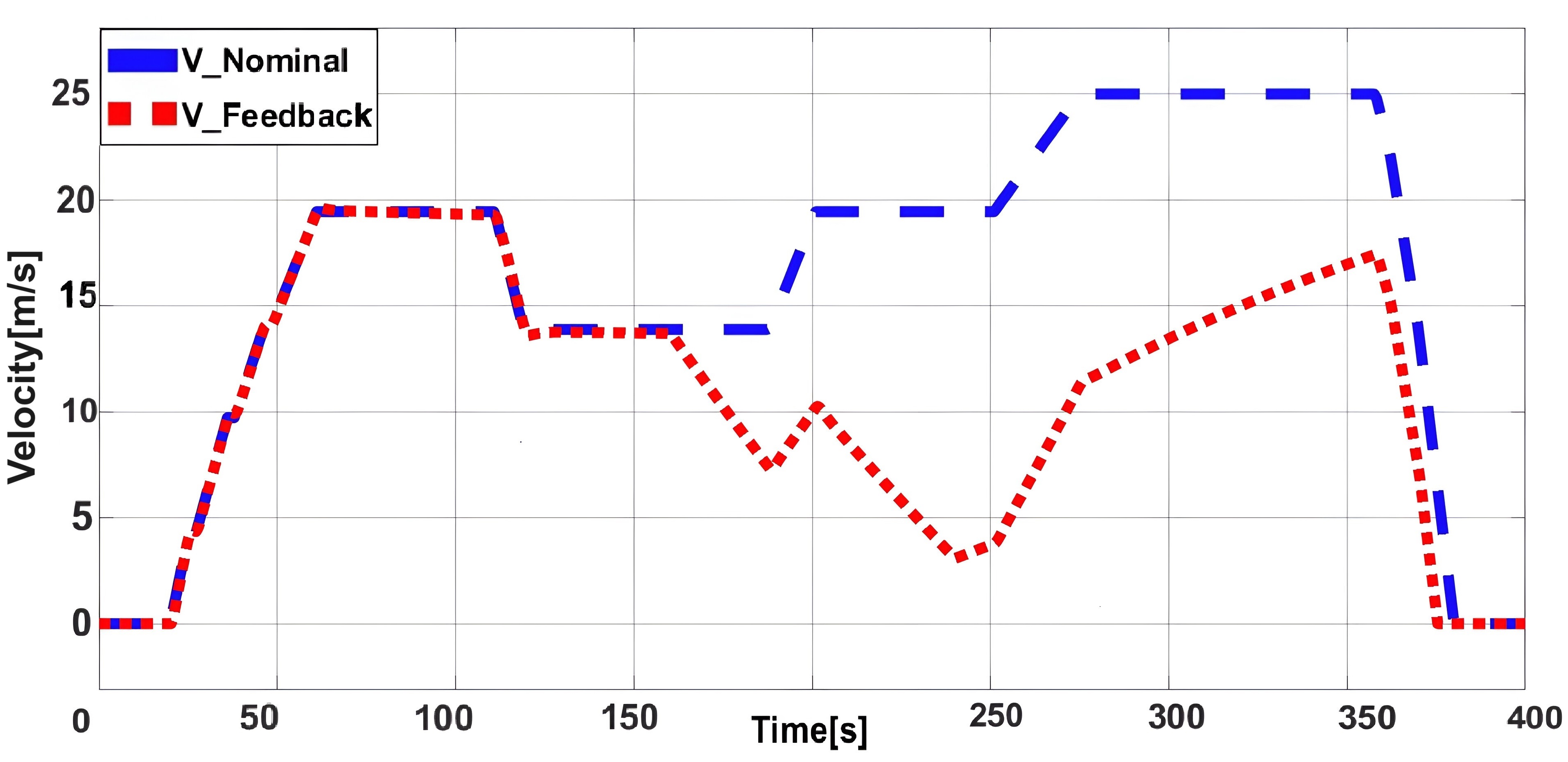}
    \caption{Velocity trajectory during torque attack and in regular driving condition while the vehicle is in Extra Urban Driving Cycle. 
    }
    \label{fig:velocity_zoomed}
\end{figure*}

Similar to the torque scenarios, both velocity signals followed the same pattern until around $t=160 s$, when the attack was executed (Figure ~\ref{fig:velocity_zoomed}). Instead of maintaining a steady velocity, the vehicle began to decelerate at this juncture, failing to adhere to the prescribed velocity trajectory. This behavior was a direct result of the imposed braking torque, which co-occurred as the vehicle attempted to accelerate according to the trajectory plan.

\subsection{Cruise Mode}
\label{subsec:cruise-mode}

\begin{figure*}[ht]
    \centering
    \includegraphics[width=0.75\textwidth]{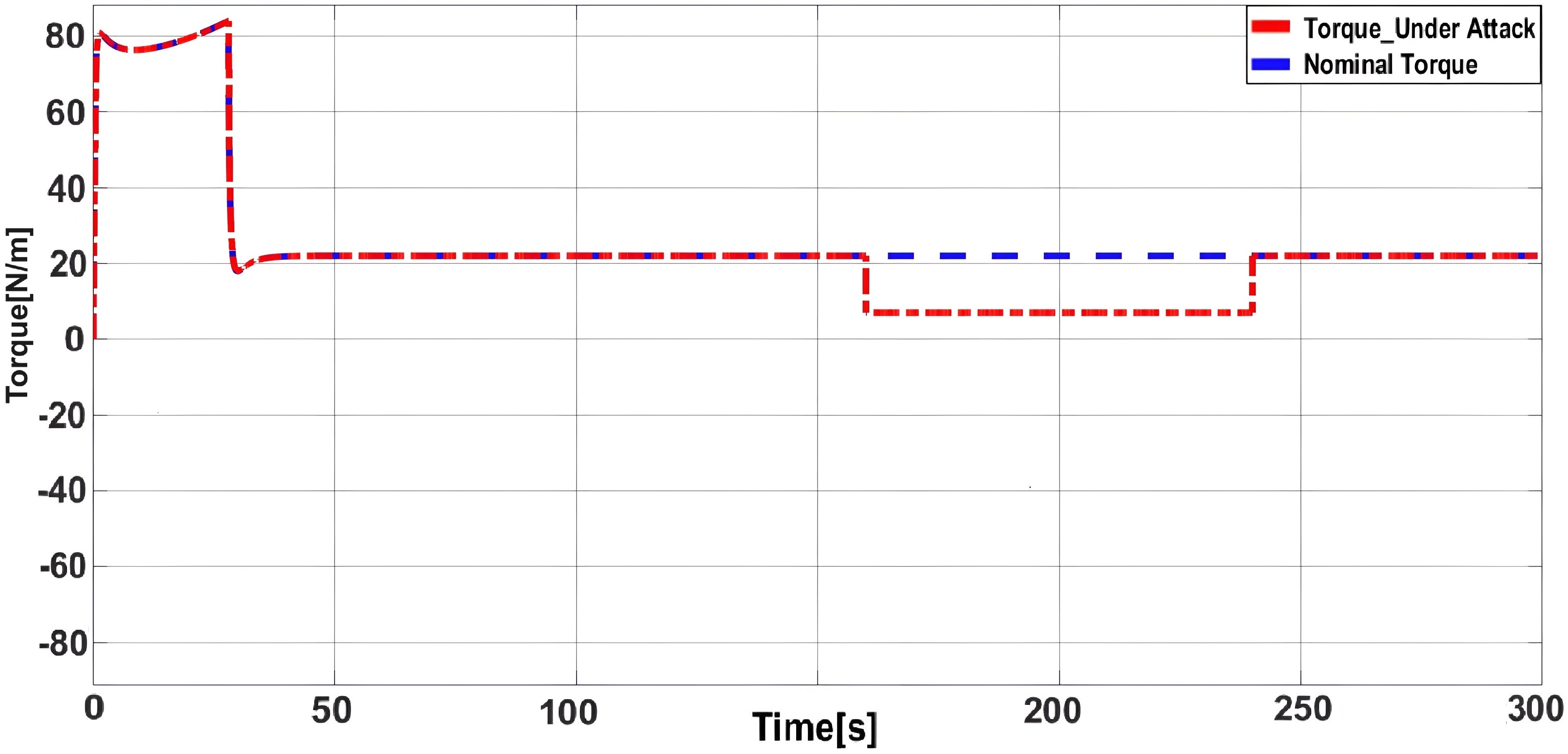}
    \caption{Torque without and with the attack while the vehicle is in Cruise Mode. 
    }
    \label{fig:torque_cruise}
\end{figure*}

The second use case simulates an attack while the vehicle was on cruise mode at 100 km/h, with the attack signal initiated at $t = 160s$ and continued until $t = 240s$. This can be seen in the torque graph comparison (Figure ~\ref{fig:torque_cruise} ), where till $t = 160s$, no attack was active, and both the non-attacked torque (represented in blue) and the attacked torque (represented in yellow) displayed identical values. At $t = 160s$, the non-attacked torque remained stable at approximately $11 Nm$; however, upon application of the attack, the torque dropped by $15 Nm$, the same magnitude as the signal created, shifting the graph downwards.

\begin{figure*}[tb]
    \centering
    \includegraphics[width=0.75\textwidth]{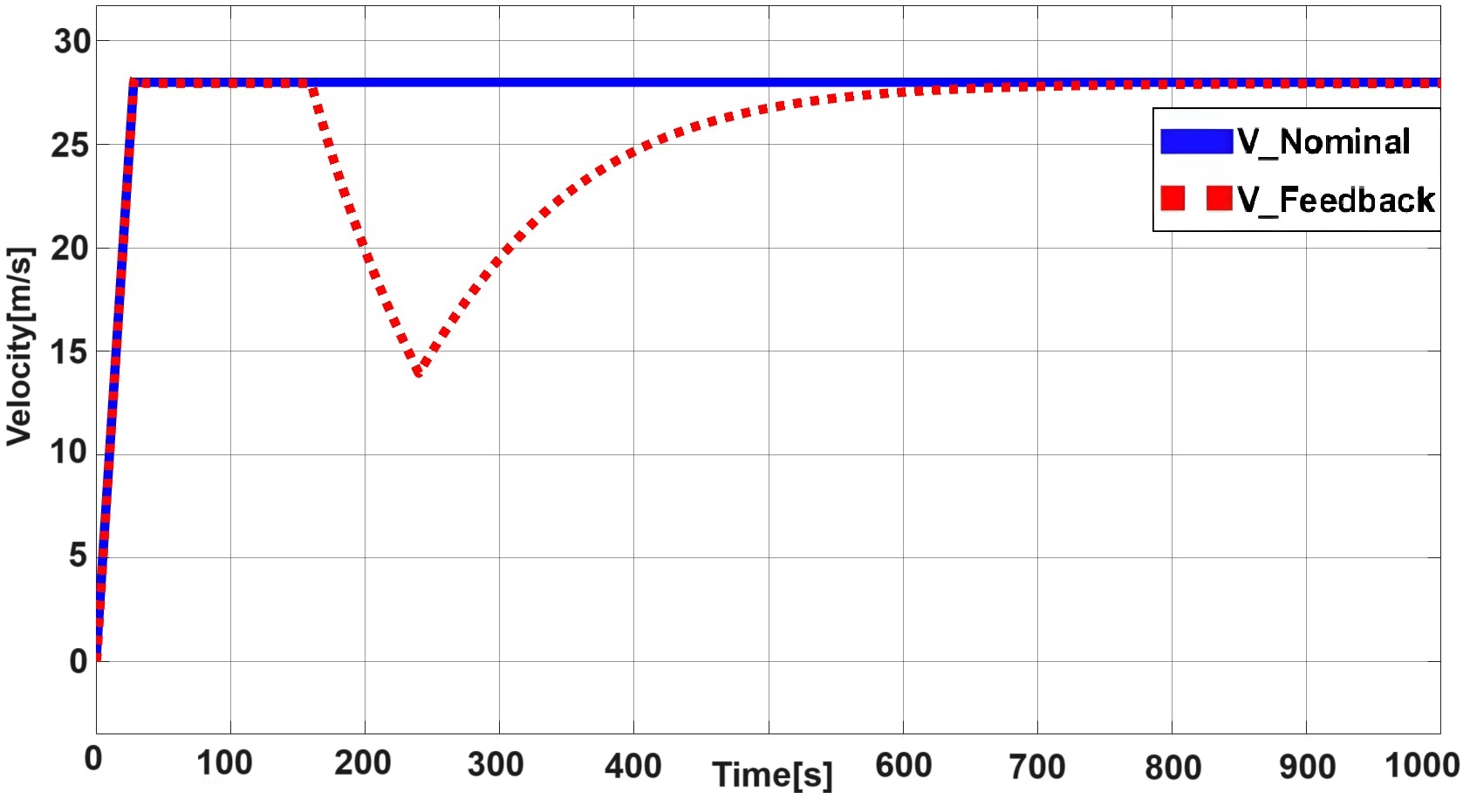}
    \caption{Velocity trajectory during torque attack and in regular driving condition while the vehicle is in Cruise Mode. 
    }
    \label{fig:velocity_cruise}
\end{figure*}

The impact of this modification is further visible in the velocity plot shown in Figure~\ref{fig:velocity_cruise}. Here, both signals showed the same trend until around $t=160s$ when the attack occurred. The car that is maintaining a constant velocity of $100Km/h$ at $t=160s$ suddenly starts to decelerate undesirably until $t=240s$ and loses its trajectory of the cruise mode. As soon as the attack was stopped, the vehicle began accelerating again, trying to reach the target velocity.

\section{Conclusion}
\label{sec:conclusion}

This paper presents a systematic approach to simulate and analyze cybersecurity attacks, mainly focusing on \gls{can} injection attacks targeting \gls{bev}. Our research addresses the rising concerns regarding the security and safety of vehicular systems in light of evolving cyber threats.

A novel simulation framework capable of generating synthetic \gls{can} traffic and simulating various attack scenarios enables the emulation of \gls{can} injection attacks on \gls{bev}, allowing direct observation of their effects on vehicle operational integrity and safety.

Future endeavors should focus on refining simulation models, exploring additional attack scenarios, and assessing the effectiveness of diverse cybersecurity measures in mitigating \gls{can} attacks.
To encourage research in this field, we release the code
related to our experiments as open-source: \url{https://github.com/smilies-polito/CARACAS}



\bibliographystyle{ieeetr}
\bibliography{biblio}

\begin{thebibliography}{10}

\bibitem{Kim:2020aa}
S.~Kim and R.~Shrestha, {\em Automotive Cyber Security: Introduction, Challenges, and Standardization}.
\newblock Springer Nature Singapore, 2021.

\bibitem{Upstream:aa}
Upstream, ``Global automotive cybersecurity report.'' upstream.auto.
\newblock \url{https://upstream.auto/reports/global-automotive-cybersecurity-report/} (accessed May 20, 2024).

\bibitem{Szydlowski:1992aa}
C.~P. Szydlowski, ``Can specification 2.0: Protocol and implementations 921603,'' in {\em SAE Technical Paper 921603}, 1992.

\bibitem{Avatefipour:2017aa}
O.~Avatefipour and H.~Malik, ``State-of-the-art survey on in-vehicle network communication \"can-bus\" security and vulnerabilities,'' {\em International Journal of Computer Science and Network}, vol.~6, pp.~720--727, 12 2017.

\bibitem{Oberti:2021aa}
F.~Oberti, E.~Sanchez, A.~Savino, F.~Parisi, and S.~di~Carlo, ``Taurum p2t: Advanced secure can-fd architecture for road vehicle,'' in {\em 2021 IEEE 27th International Symposium on On-Line Testing and Robust System Design (IOLTS)}, pp.~1--7, 2021.

\bibitem{Oberti:2024aa}
F.~Oberti, A.~Savino, E.~Sanchez, P.~Casasso, F.~Parisi, and S.~D. Carlo, ``{CAN-MM: Multiplexed Message Authentication Code for Controller Area Network Message Authentication in Road Vehicles},'' {\em IEEE Transactions on Vehicular Technology}, pp.~1--13, 2024.

\bibitem{Bi:2023aa}
Z.~Bi, G.~Xu, C.~Wang, G.~Xu, and S.~Zhang, ``A method for translating automotive body-related can messages based on labeled bits,'' {\em Applied Sciences}, vol.~13, no.~3, 2023.

\bibitem{stelvio}
{Stelvio Forum}, ``Peace of mind car hacking relay attack,'' 2023.

\bibitem{Jo:2022aa}
H.~J. Jo and W.~Choi, ``A survey of attacks on controller area networks and corresponding countermeasures,'' {\em IEEE Transactions on Intelligent Transportation Systems}, vol.~23, no.~7, pp.~6123--6141, 2022.

\bibitem{Han:2018aa}
M.~L. Han, B.~I. Kwak, and H.~K. Kim, ``Anomaly intrusion detection method for vehicular networks based on survival analysis,'' {\em Vehicular Communications}, vol.~14, pp.~52--63, 2018.

\bibitem{IoTProxy2024}
D.~Canavese, L.~Mannella, L.~Regano, and C.~Basile, ``{Security at the Edge for Resource-Limited IoT Devices},'' {\em Sensors}, vol.~24, no.~2, 2024.

\bibitem{Chenet:2024aa}
C.~P. Chenet, A.~Savino, and S.~Di~Carlo, ``A survey on hardware-based malware detection approaches,'' {\em IEEE Access}, vol.~12, pp.~54115--54128, 2024.

\bibitem{Song:2016aa}
H.~M. Song, H.~R. Kim, and H.~K. Kim, ``{Intrusion detection system based on the analysis of time intervals of CAN messages for in-vehicle network},'' in {\em 2016 International Conference on Information Networking (ICOIN)}, pp.~63--68, 2016.

\bibitem{Lokman:2019aa}
S.-F. Lokman, A.~T. Othman, and M.-H. Abu-Bakar, ``{Intrusion detection system for automotive Controller Area Network (CAN) bus system: a review},'' {\em EURASIP Journal on Wireless Communications and Networking}, vol.~2019, July 2019.

\bibitem{Seo:2019aa}
E.~Seo, H.~M. Song, and H.~K. Kim, ``{GIDS: GAN based Intrusion Detection System for In-Vehicle Network},'' {\em In 2018 16th Annual Conference on Privacy, Security and Trust (PST), pp. 1-6. IEEE, 2018}, 07 2019.

\bibitem{Lee:2017aa}
H.~Lee, S.~H. Jeong, and H.~K. Kim, ``Otids: A novel intrusion detection system for in-vehicle network by using remote frame,'' in {\em 2017 15th Annual Conference on Privacy, Security and Trust (PST)}, IEEE, Aug. 2017.

\bibitem{Verma:2020aa}
M.~E. Verma, R.~A. Bridges, M.~D. Iannacone, S.~C. Hollifield, P.~Moriano, S.~C. Hespeler, B.~Kay, and F.~L. Combs, ``{A Comprehensive Guide to CAN IDS Data \& Introduction of the ROAD Dataset},'' {\em PLoS one 19, no. 1 (2024): e0296879}, 12 2020.

\bibitem{MATLAB:2024aa}
MATLAB, ``Simscape.'' \url{https://it.mathworks.com/products/matlab.html}, 2024.
\newblock \url{https://mathworks.com/products/simscape.html} (accessed May 20, 2024).

\bibitem{Documentation:2020aa}
S.~Documentation, ``Simulation and model-based design,'' 2020.

\bibitem{Hoppe:2011aa}
T.~Hoppe, S.~Kiltz, and J.~Dittmann, ``Security threats to automotive can networks - practical examples and selected short-term countermeasures,'' {\em Reliability Engineering \& System Safety}, vol.~96, no.~1, pp.~11--25, 2011.
\newblock Special Issue on Safecomp 2008.

\bibitem{Gazdag:2023aa}
A.~Gazdag, R.~Ferenc, and L.~Butty{\~A}{!'}n, ``{CrySyS dataset of CAN traffic logs containing fabrication and masquerade attacks},'' {\em Scientific Data}, vol.~10, Dec. 2023.

\bibitem{Dupont:2019aa}
G.~Dupont, A.~Lekidis, J.~J. den Hartog, and S.~S. Etalle, ``Automotive controller area network (can) bus intrusion dataset v2,'' 2019.

\bibitem{Hanselmann:2020aa}
M.~Hanselmann, T.~Strauss, K.~Dormann, and H.~Ulmer, ``{CANet: An Unsupervised Intrusion Detection System for High Dimensional CAN Bus Data},'' {\em IEEE Access}, vol.~8, pp.~58194--58205, 2020.

\bibitem{hacktool}
{CAN Hacker}, ``{CAN Hacker: Tools and Resources for CAN Bus Hacking},'' 2023.
\newblock Accessed: 2023-05-05.

\bibitem{Vector-Informatik-GmbH:2024aa}
{Vector Informatik GmbH}, ``Candb++ editor,'' 2024.

\end{thebibliography}

\vspace{12pt}
\end{document}